# Decision-making Strategy on Highway for Autonomous Vehicles using Deep Reinforcement Learning

**Jiangdong Liao[1], Teng Liu[2], Xiaolin Tang[2], Xingyu Mu[2], Bing Huang[2], and Dongpu Cao[3]**
[1]School of Mathematics and Statistics, Yangtze Normal University, Fuling, 408100, China
[2]College of Automotive Engineering, Chongqing University, Chongqing, 400044, China
[3]Department of Mechanical and Mechatronics Engineering, University of Waterloo, N2L 3G1, Canada

Corresponding author: Teng Liu (e-mail: tengliu17@gmail.com), Xiaolin Tang (e-mail: tangxl0923@cqu.edu.cn).

This work was in part supported by the State Key Laboratory of Mechanical System and Vibration (Grant No. MSV202016).

**ABSTRACT** Autonomous driving is a promising technology to reduce traffic accidents and improve driving efficiency. In this work, a deep reinforcement learning (DRL)-enabled decision-making policy is constructed for autonomous vehicles to address the overtaking behaviors on the highway. First, a highway driving environment is founded, wherein the ego vehicle aims to pass through the surrounding vehicles with an efficient and safe maneuver. A hierarchical control framework is presented to control these vehicles, which indicates the upper-level manages the driving decisions, and the lower-level cares about the supervision of vehicle speed and acceleration. Then, the particular DRL method named dueling deep Q-network (DDQN) algorithm is applied to derive the highway decision-making strategy. The exhaustive calculative procedures of deep Q-network and DDQN algorithms are discussed and compared. Finally, a series of estimation simulation experiments are conducted to evaluate the effectiveness of the proposed highway decision-making policy. The advantages of the proposed framework in convergence rate and control performance are illuminated. Simulation results reveal that the DDQN-based overtaking policy could accomplish highway driving tasks efficiently and safely.

## I. INTRODUCTION

Autonomous driving (AD) enables the vehicle to engage different driving missions without a human driver [1, 2]. Motivated by the enormous potentials of artificial intelligence (AI), autonomous vehicles or automated vehicles have become one of the research hotspots all over the world [3]. Many automobile manufacturers, such as Toyota, Tesla, Ford, Audi, Waymo, Mercedes-Benz, General Motors, and so on are developing their own autonomous cars and achieving tremendous progress. Meanwhile, automotive researchers are paying attention to overcome the essential technologies to build automated cars with full automation [4].

Four significant modules are contained in autonomous vehicles, which are perception, decision-making, planning, and control [5]. Perception indicates the autonomous vehicles know the information about the driving environments based on the functions of a variety of sensors, such as radar, lidar, global positioning system (GPS), et al. Decision-making controller manages the driving behaviors of the vehicles, and these behaviors include acceleration, braking, lane-changing, lane-keep and so on [6]. Planning function helps the automated cars find the reasonable running trajectories from one point to another. Finally, the control module would command the onboard powertrain components to operate accurately to finish the driving maneuvers and follow the planning path. According to the intelligent degrees of these mentioned modules, the AD is classified into six levels, from L0 to L5 [7].

Decision-making strategy is regarded as the human brain and is extremely important in autonomous vehicles [8]. This policy is often generated by the manual rules based on human driving experiences or imitated manipulation learned from supervised learning approaches. For example, Song et al. applied a continuous hidden Markov chain to predict the motion intention of the surrounding vehicles. Then, a partially observable Markov decision process (POMDP) is used to construct the general decision-making framework [9]. The authors in [10] developed an advanced ability to make appropriate decisions in the city road traffic situations. The presented decision-making policy is multiple criteria, which helps the city cars make feasible choices in different conditions. In Ref. [11], Nie et al. discussed the lane-changing decision-making strategy for connected automated cars. The related model is combining the cooperative car-following models and candidate decision generation module. Furthermore, the authors in [12] mentioned the thought of a human-like driving system. It could adjust the driving decisions by considering the driving demand of human drivers.

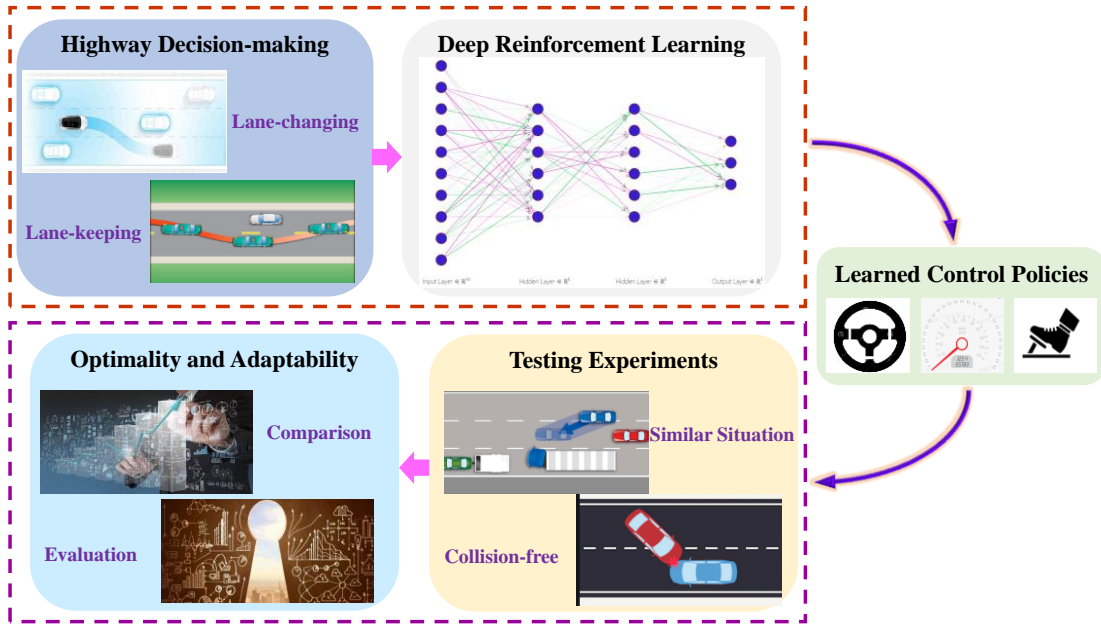

FIGURE 1. The constructed deep reinforcement learning-abled highway overtaking driving policy for autonomous vehicles.

Deep reinforcement learning (DRL) techniques are taken as a powerful tool to deal with the long sequential decision-making problems [13]. In recent years, many attempts have been implemented to study DRL-based autonomous driving topics. For example, Duan et al. built a hierarchical structure to learn the decision-making policy via the reinforcement learning (RL) method [14]. The pro of this work is independent of the historical labeled driving data. Ref. [15, 16] utilized DRL approaches to handle the collision avoidance and path following problems for automated vehicles. The relevant control performance is better than the conventional RL methods in these two findings. Furthermore, the authors in [17] considered not only path planning but also the fuel consumption for autonomous vehicles. The related algorithm is deep Q-learning (DQL), and it was proven to accomplish these two-driving missions suitably. Han et al. employed the DQL algorithm to decide the lane change or lane keep for connected autonomous cars, in which the information of the nearby vehicles is treated as feedback knowledge from the network [18]. The resulted policy is able to promote traffic flow and driving comfort. However, the common DRL methods are unable to address the highway overtaking problems because of the continuous action space and large state space [19].

In this work, a DRL enabled highway overtaking driving policy is constructed for autonomous vehicles. The proposed decision-making strategy is evaluated and estimated to be adaptive to other complicated scenarios, as depicted in Fig. 1. First, the studied driving environment is founded on the highway, wherein an ego vehicle aims to run through a particular driving scenario efficiently and safely. Then, a hierarchical control structure is shown to manipulate the lateral and longitudinal motions of the ego and surrounding vehicles. Furthermore, the special DRL algorithm called dueling deep Q-network (DDQN) is derived and utilized to obtain the highway decision-making strategy. The DQL and DDQN algorithms are compared and analyzed theoretically. Finally, the performance of the proposed control framework is discussed via executing a series of simulation experiments. Simulation results reveal that the DDQN-based overtaking policy could accomplish highway driving tasks efficiently and safely.

The main contributions and innovations of this work can be cast into three perspectives: 1) an adaptive and optimal DRL-based highway overtaking strategy is proposed for automated vehicles; 2) the dueling deep Q-network (DDQN) algorithm is leveraged to address the large state space of the decision-making problem; 3) the convergence rate and control optimization of the derived decision-making policy are demonstrated by multiple designed experiments.

This following organization of this article is given as follows: the highway driving environment and the control modules of the ego and surrounding vehicles are described in Section II. The DQL and DDQN algorithms are defined in Section III, in which the parameters of the RL framework are discussed in detail. Section IV shows the relevant results of a series of simulation experiments. Finally, the conclusion is conducted in Section V.

## II. DRIVING ENVIRONMENT AND CONTROL MODULE

In this section, the studied driving scenario on the highway is introduced. Without loss of generality, a three-lane freeway environment is constructed. Furthermore, a hierarchical motion controller is described to manage the lateral and longitudinal movements of the ego and surrounding vehicles. The upper-level contains two models, which are intelligent driver model (IDM) and minimize overall braking induced by lane changes (MOBIL) [20]. The lower-level focuses on regulating vehicle velocity and acceleration.

### A. HIGHWAY DRIVING SCENARIO

Decision-making in autonomous driving means selecting a sequence of reasonable driving behaviors to achieve special driving missions. On the highway, these behaviors involve lane- changing, lane-keeping, acceleration, braking. The main objectives are avoiding collisions, running efficiently, and driving on the preferred lane. Accelerating and surpassing other vehicles is a typical driving behavior called overtaking.

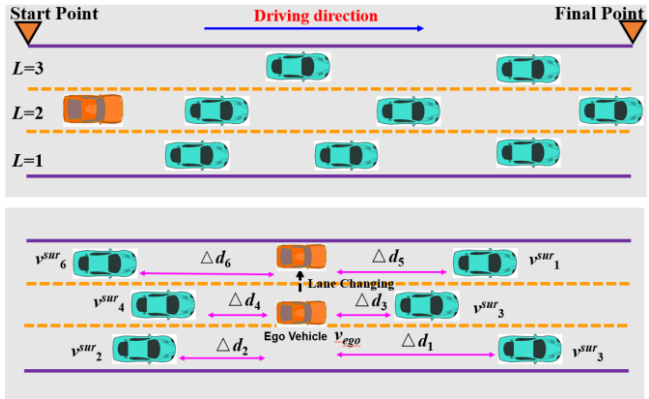

**FIGURE 2.** Highway driving environment for decision-making problem with three lanes.

This work discusses the decision-making problem on the highway for autonomous vehicles, and the research driving scenario is depicted in Fig. 2. The orange vehicle is the ego vehicle, and other green cars are named as surrounding vehicles. There are three lanes in the driving environment, and the derived decision-making policy in this paper is easily generalized to different situations. The ego vehicle would be initialized in the middle lane at a random speed.

The objective of the ego vehicle is to run from the starting point to end point as soon as possible with cashing the surrounding vehicles. Hence, this goal is interpreted as efficiency and safety. The initial velocity and position of the surrounding vehicles are designed randomly. It implies the driving scenario consists of uncertainties as to the actual driving. Furthermore, to imitate the real conditions, the ego vehicle prefers to stay on lane 1 ($L$=1), and it can overtake other vehicles from the right or left sides.

At the beginning of this driving task, all the surrounding vehicles located in front of the ego vehicle. In each lane, the number of surrounding vehicles is $M$, which indicates there are $3M$ nearby cars in this situation. Two conditions would interrupt the ego vehicle, which is crashing other vehicles or reaching the destination. The procedure of running from the starting point to the ending point is called as one episode in this work.

Without loss of generality, the parameters of the driving scenario are settled as follows: the original speed of the ego vehicle is chosen from [23, 25] m/s, its maximum speed is 40 m/s, the length and width of all vehicles are 5m and 2m. The duration of one episode is 100s, and the simulation frequency is 20 Hz. The initial velocity of the surrounding vehicles is randomly chosen from [20, 23] m/s, and their behaviors are manipulated by IDM and MOBIL. The next section will discuss these two models in detail.

### B. VEHICLE BEHAVIOR CONTROLLER

The movements of all the vehicles in the highway environments are mastered by a hierarchical control framework, as shown in Fig. 3. The upper-level applied IDM and MOBIL to manage the vehicle behaviors, and the lower-level aims to enable the ego vehicle to track a given target speed and follow a target lane. In this work, the DRL method is used to control the ego vehicle. The reference model implies that the ego vehicle is controlled by the bi-level structure in Fig. 3, which is taken as a benchmark to evaluate the DRL-based decision-making strategy.

IDM in the upper-level is a prevalent microscopic model [21] to realize car-following and collision-free. In the adaptive cruise controller of automated cars, the longitudinal behavior is usually decided by IDM. In general, the longitudinal acceleration is IDM is determined as [22]:

$$a = a_{\max} \cdot [1 - (\frac{v}{v_{tar}})^\delta - (\frac{d_{tar}}{\Delta d})^2] \quad (1)$$

where $v$ and $a$ is the current vehicle speed and acceleration. $a_{\max}$ is the maximum acceleration, $\triangle d$ is the distance to the front car and $\delta$ is named as constant acceleration parameter. $v_{tar}$ and $d_{tar}$ are the target velocity and distance, and the desired speed is achieved by the $a_{max}$ and $d_{tar}$. In IDM, the expected distance $d_{tar}$ is affected by the front vehicle and is calculated as follows:

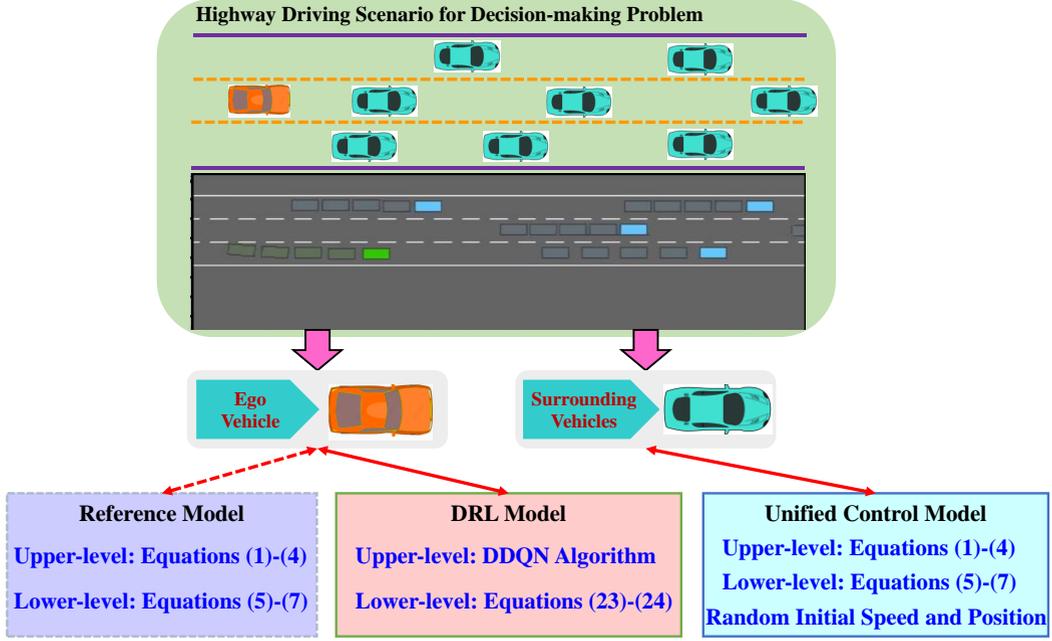

**FIGURE 3.** The hierarchical control framework discussed in this work for the ego vehicle and surrounding vehicles.

$$d_{tar} = d_0 + Tv + \frac{v\Delta v}{2\sqrt{a_{max}b}} \qquad (2)$$

where $d_0$ is the predefined minimum relative distance, $T$ is the expected time interval for safety goal, $\triangle v$ is the relative speed between two vehicles, and $b$ is the deceleration rate according to the comfortable purpose.

In IDM, the relative speed and distance are defined a priori to induce the vehicle velocity and acceleration at each time step. The default configuration is introduced as following: the maximum acceleration $a_{max}$ is 6 m/s², acceleration argument $\delta$ is 4, desired time gap $T$ is 1.5 s, comfortable deceleration rate $b$ is -5 m/s², and minimum relative distance $d_0$ is 10m.

Since the IDM is utilized to determine the longitudinal behavior, the MOBIL is employed to make the lateral lane change decisions [23]. MOBIL states that lane-changing behaviors should be observed by two restrictions, which are safety criterion and incentive condition. These constraints are related to the ego vehicle and its two followers (before changing and after changing). Assuming $a^{old}_i$ and $a^{old}_j$ are the accelerations of these followers before changing, and $a^{new}_i$ and $a^{new}_j$ are the accelerations after changing.

The safety criterion requires the follower in the desired lane (after changing) to limit its acceleration to avoid a collision. The mathematic expression is shown as:

$$a^{new}_j \geq -b_{safe} \qquad (3)$$

where $b_{safe}$ is the maximum braking imposed to the follower in the lane-changing behavior. By following (3), collision and accidents could be avoided effectively.

The incentive condition is imposed on the ego vehicle and its followers by an acceleration threshold $a_{th}$:

$$a^{new}_e - a^{old}_e + z[(a^{new}_i - a^{old}_i) + (a^{new}_j - a^{old}_j)] > a_{th} \qquad (4)$$

where $z$ is named as politeness coefficient to determine the effect degree of the followers in the lane-changing behaviors. This incentive condition means the desired lane should be safer than the old lane. For application, the parameters in MOBIL are defined as follows: the politeness factor z is 0.001, safe deceleration limit $b_{safe}$ is 2 m/s², and acceleration threshold $a_{th}$ is 0.2 m/s². After deciding the longitudinal and lateral behaviors in the upper-level, the lower-level is applied to follow the target speed and lane.

### C. VEHICLE MOTION CONTROLLER

In the lower-level, the motions of the vehicles in the longitudinal and lateral direction are controlled. The former regulates the acceleration by a proportional controller as:

$$a = K_p(v_{tar} - v) \qquad (5)$$

where $K_p$ is the proportional gain.

In the lateral direction, the controller deals with the position and heading of the vehicle with a simple proportional-derivative action. The position indicates the lateral speed $v_{lat}$ of the vehicle is computed as follows:

$$v_{lat} = -K_{p,lat}\Delta_{lat} \qquad (6)$$

where $K_{p,lat}$ is named as position gain, $\triangle_{lat}$ is the lateral position of the vehicle with respect to the center-line of the lane. Then, the heading control is related to the yaw rate command φ as:

$$\dot{\varphi} = K_{p,\varphi}(\varphi_{tar} - \varphi) \quad (7)$$

where $\varphi_{tar}$ is the target heading angle to follow the desired lane and $K_{p,lat}$ is the heading gain.

Hence, the movements of the surrounding vehicles are achieved by the bi-level control framework in Fig. 3. The position, speed, and acceleration of these vehicles are assumed to be known to the ego vehicle. This limitation propels the ego vehicle to learn how to drive in the scenario via the trial-and-error procedure. In the next section, the DRL approach is introduced and established to realize this learning process and derive the highway decision-making policy.

### III. DRL METHODOLOGY

This section introduces the RL method and exhibits the special DRL algorithms. The interaction in RL between the agent and the environment is first explained. Then, the DQL algorithm that incorporates the neural network and Q-learning algorithm is formulated. Finally, a dueling network is constructed in a DQL algorithm to reconstitute the output layer of the neural network, and thus raise the DDQN method.

#### A. RL CONCEPT

RL approach describes the process that an intelligent agent interacts with its environment. It is powerful and useful to solve sequential decision-making problems. The goal of the agent is to search an optimal sequence of control actions based on feedback from the environment. Owing to its characteristics of self-evaluation and self-promotion, RL is widely used in many research fields [24-28].

In the decision-making problem on the highway, the agent and environment are the ego vehicle and surrounding vehicles (including the driving conditions), respectively. The Markov decision processes (MDPs) is embedded in the Markov property that the next state variable is only concerned with the current state and action [29]. This MDP is often utilized to represent the RL interaction as a tuple (*S*, *A*, *P*, *R*, *γ*), in which *S* and *A* are the state and control sets. *P* and *R* are the significant elements of the environments in RL, and they mean the transition and reward model, respectively. In RL, the current action would influence the immediate and future rewards synchronously. Hence, *γ* is a discount factor to balance these two parts of rewards.

To represent the list of future rewards, the accumulated reward $R_t$ is defined as follows:

$$R_t = \sum_{t}^{\infty} \gamma^t \cdot r_t \quad (8)$$

where *t* is the time instant, and $r_t$ is the relevant reward. To record the worth of the state *s* and state-action pair (*s*, *a*), two value function as expressed by the accumulated reward as:

$$V^\pi(s_t) \doteq E_\pi[R_t \mid s_t, \pi] \quad (9)$$
$$Q^\pi(s_t, a_t) \doteq E_\pi[R_t \mid s_t, a_t, \pi] \quad (10)$$

where *π* is called as control action policy, *V* is state value function, and *Q* is the state-action function (called Q table for short). To be updated easily, the state-action function is usually rewritten as the recursive form:

$$Q^\pi(s_t, a_t) = E_\pi[r_t + \gamma \max_{a_{t+1}} Q^\pi(s_{t+1}, a_{t+1})] \quad (11)$$

Finally, the optimal control action with respect to the control policy *π* is determined by the state-action function:

$$\pi(s_t) = \arg\max_{a_t} Q(s_t, a_t) \quad (12)$$

Therefore, the essence of different RL algorithms is updating the state-action function $Q(s, a)$ in various ways. According to the style of updating rules, the RL algorithms could have diverse classifications, such as model-based and model-free, policy-based and value-based, temporal-difference (TD), and Monte-Carlo (MC) [30].

#### B. DEEP Q NETWORK

Deep Q network (DQN) is first presented to play the Atari games in [31]. It synthesizes the strengths of deep learning (neural network) and Q-learning to obtain the new state-value function. In the common Q-learning, the updating rule of this function is narrated as follows:

$$Q(s, a) \leftarrow Q(s, a) + \alpha[r + \gamma \max_{a'} Q(s', a') - Q(s, a)] \quad (13)$$

where $\alpha \in [0, 1]$ is named as a learning rate to trade-off the old and new learned experiences from the environment. *s´* and *a´* are the state and action at the next time step.

The common Q-learning is unable to handle the problem with a large space of state variable, because it needs an enormous time to obtain the mutable Q table. Thus, in DQN, a neural network is employed to approximate the Q table as $Q(s, a; \theta)$. For the neural network, the inputs are the arrays of state variables and control actions, and the output is the state-value function [31].

To measure the discrepancy between the approximated and actual Q table in DQN, the loss function is introduced like the following expression:

$$L(\theta) = E[\sum_{t=1}^{N}(y_t - Q(s, a; \theta))^2] \quad (14)$$

where

$$y_t = r_t + \gamma \max_{a'} Q(s', a'; \theta') \quad (15)$$

As can be seen, there are two parameters (*θ* and *θ´*) of the neural network, which delegate two networks in DQN. These networks are prediction and target networks. The former is applied to estimate the current control action, and the latter aims to generate the target value. In general, the target network would copy the parameters from the prediction network every certain number of time steps. By doing this, the target Q table will converge to predict one to some extent to remit the network *instability*.

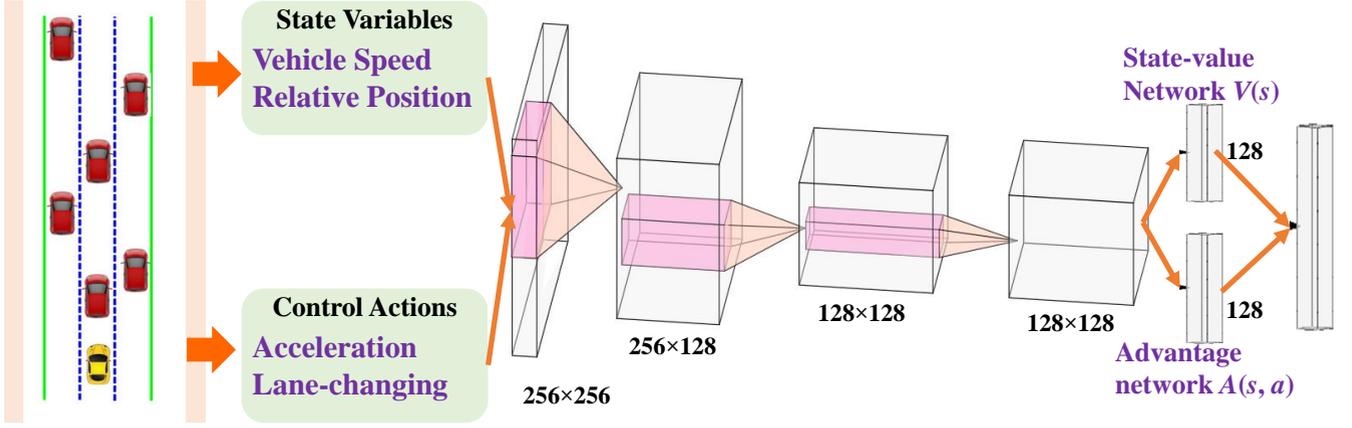

**FIGURE 4.** The dueling network combined with state-value network and advantage network for Q table updating.

In DQN, the online neural network is updated by gradient descent as follows:

$$\nabla_\theta L(\theta) = E[(y_i - Q(s,a;\theta))\nabla_\theta Q(s,a;\theta)] \quad (16)$$

This operation makes the DQN as an off-policy algorithm, and the states and rewards are acquired by a special criterion. This rule is known as epsilon greedy, which indicates that the agent executes the exploration (choose a random action) with probability $\varepsilon$, and makes exploitation (use the current best action) with probability $1-\varepsilon$.

### C. DUELING DQN ALGORITHM

In some RL problems, the selection of current control action may not cause negative results, apparently. For example, in the highway environment, many actions would not lead to the collision. However, these choices may indirectly result in bad rewards afterward [32]. Motivated by this insight, a dueling network is proposed in this work to estimate the worth of the control actions at each step. A new neural network is constructed to approximate the Q table in the highway decision-making problem, as shown in Fig. 4.

Two streams of fully connected layers are used to estimate the state-value function $V(s)$ and the advantage function $A(s, a)$ of each action. Therefore, the state-action function (Q table) is constituted as follows:

$$Q^\pi(s,a) = A^\pi(s,a) + V^\pi(s) \quad (17)$$

It is obvious that the output of this new dueling network is also a Q table, and thus the neural network used in DQN can also be employed to approximate this Q table. The network with two parameters is computed as:

$$Q^\pi(s,a;\theta) = V^\pi(s;\theta_1) + A^\pi(s,a;\theta_2) \quad (18)$$

where $\theta_1$ and $\theta_2$ are the parameters of state-value function and advantage function, respectively.

To update the Q table in DDQN and achieve the optimal control action, (18) is reformulated as follows:

$$Q^\pi(s,a;\theta) = V^\pi(s;\theta_1) + (A^\pi(s,a;\theta_2) - \max_{a'} A^\pi(s,a';\theta_2)) \quad (19)$$

$$a^* = \arg\max_{a'} Q(s,a';\theta) = \arg\max_{a'} A(s,a';\theta_2) \quad (20)$$

It can be decerned that the input-output interfaces in DDQN and DQN are the same. Hence, the gradient descent in (16) is capable of being recycled to train the Q table in this work.

### D. VARIABLES SPECIFICATION

To derive the DDQN-based decision-making strategy, the preliminaries are initialized as follows, and the calculative procedure is easily transformed into an analogous driving environment. The control actions are the longitudinal and lateral accelerations ($a_1$ and $a_2$) with the units m/s² and rad:

$$a_1 \in [-5, 5] \text{ m/s}^2 \quad (21)$$

$$a_2 \in [-\pi/4, \pi/4] \text{ rad} \quad (22)$$

It is noticed that when these two accelerations are zeros, the ego vehicle adopts an idling control.

After obtaining the acceleration actions, the speed and position of the vehicle can be computed as follows:

$$\begin{cases} v_1^{t+1} = v_1^t + a_1 \cdot \Delta t \\ v_2^{t+1} = v_2^t + a_2 \cdot \Delta t \end{cases} \quad (23)$$

$$\begin{cases} d_1^t = v_1^t \cdot \Delta t + \frac{1}{2} \cdot a_1 \cdot \Delta t^2 \\ d_2^t = v_2^t \cdot \Delta t + \frac{1}{2} \cdot a_2 \cdot \Delta t^2 \end{cases} \quad (24)$$

where $v_1$, $v_2$ are the longitudinal and lateral speed of the vehicle, respectively, similar to the $d_1$ and $d_2$. The policy frequency is 1 Hz, which indicates the time interval $\triangle t$ is 1 second. It should be noticed that (23) and (24) are feasible for the ego vehicle and surrounding vehicles simultaneously, and these expressions are considered as the transition model $P$ in RL. Then, the state variables are defined as the relative speed and distance between the ego and nearby cars:

$$\Delta d_t = \left| d_t^{ego} - d_t^{sur} \right| \quad (25)$$

$$\Delta v_t = \left| v_t^{ego} - v_t^{sur} \right| \quad (26)$$

where the superscript *ego* and *sur* represent the ego vehicle and surrounding vehicles, respectively.

Finally, the reward model $R$ is constituted by the optimal control objectives, which are avoiding collision, running as fast as possible, and trying the driving on lane 1 ($L$=1). To bring this insight to fruition, the instantaneous reward function is defined as following:

$$r_t = -1 \cdot collision - 0.1 * (v_{ego}^t - v_{ego}^{max})^2 - 0.4 * (L-1)^2 \quad (27)$$

where $collision \in \{0, 1\}$ and the goal of the DDQN-based highway decision-making strategy is maximizing the cumulative rewards.

The proposed decision-making control policy is trained and evaluated in the simulation environment based on the OpenAI gym Python toolkit [31]. The numbers of lanes and surrounding vehicles are 3 and 30. The discount factor $\gamma$ and learning rate $\alpha$ are 0.8 and 0.2. The layers of the value network and advantage network are both 128. The value of $\varepsilon$ decreases from 1 to 0.05 with the time step 6000. The training episode in different DRL approaches are 2000. The next section discusses the effectiveness of the presented decision-making strategy for autonomous vehicles.

## IV. RESULTS AND EVALUATION

In this section, the proposed highway decision-making policy is estimated by comparing it with the benchmark methods. These techniques are the reference model in Fig. 3 and the common DQN in Section III.B. The optimality is analyzed by conducting a comparison of these three methods. Furthermore, the adaptability of the presented approach is verified by implementing the trained model into a similar highway driving scenario.

### A. OPTIMALITY EVALUATION
The reference model, DQN, and DDQN are compared in this subsection. All of them adopted a hierarchical control framework. The lower-levels are the same and utilize (5)-(7) to regulate the acceleration, position, and heading. The upper-levels are different, which are IDM and MOBIL, DQN algorithm in Section III.B, and DDQN algorithm in Section III.C. The default parameters are the same in DQN and DDQN.

Fig. 5 depicts the normalized average rewards of these three methods. Based on the definition of reward function in (27), higher reward indicates driving on the preferred lane with a more efficient maneuver. It is obvious that the training stability and learning speed of DDQN are better than the other two approaches. Besides, after about 500 episodes, the reward in DDQN is greater than the other two approaches, and it keeps this momentum all the time. And they are both better than the reference model. It is mainly caused by the advantage network in DDQN. This network could assess the worth of the chosen action at each step, which helps the ego vehicle to find a better decision-making policy fleetly.

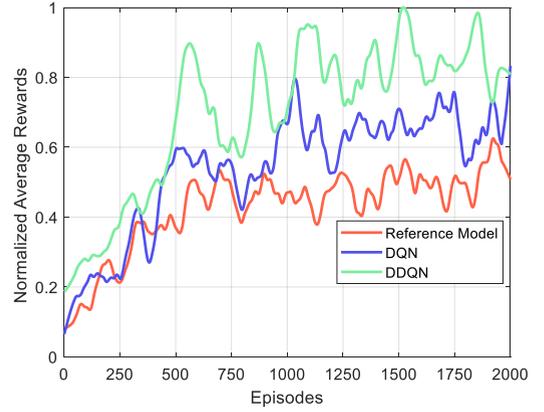

**FIGURE 5.** Average reward variation in three compared methods: the reference model, DQN and DDQN.

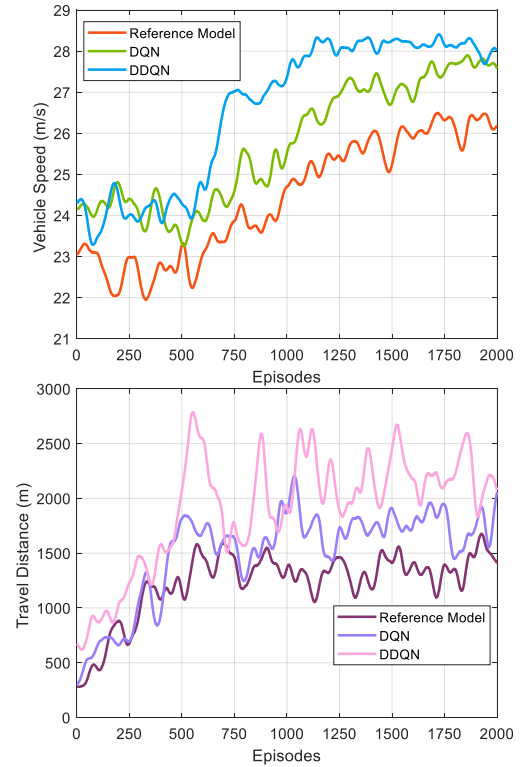

**FIGURE 6.** Vehicle speed and traveling distance of the ego vehicle in each episode of these compared techniques.

To observe the trajectories of state variables in this work, Fig. 6 shows the average vehicle speed and traveling distance in these three compared techniques. They are all trained by 2000 episodes. A higher average implies that the ego vehicle could run through the driving scenario faster and achieve greater cumulative rewards. The traveling distance means the ego vehicle could drive longer without collision. These results directly reflect safety and efficiency demand. The noticeable differences are able to certify the optimality of the proposed algorithm.

As the ego vehicle is not willing to crash the surrounding vehicles, the collision conditions of these three control cases are described in Fig. 7, wherein collision has two values (collision = 0 or 1). It can be noticed that the DDQN, DQN, and reference model-enabled agents could avoid a collision after 1300, 1700, 1950 episodes, respectively. This appearance can also prove that the DDQN-based agent is more intelligent other two agents. As the safety claim is the first concern for the actual application of automated driving, the learned decision-making model based on DDQN is more promising to be employed in real-world environments.

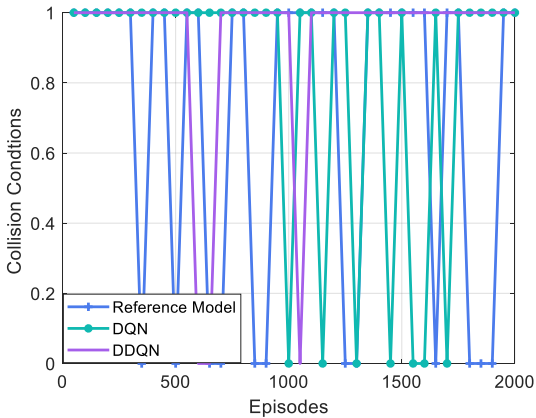

**FIGURE 7.** Collision conditions of the ego vehicle in each compared method: collision=0, the ego vehicle does not crash other vehicles; collision=1, the ego vehicle crashes other vehicles.

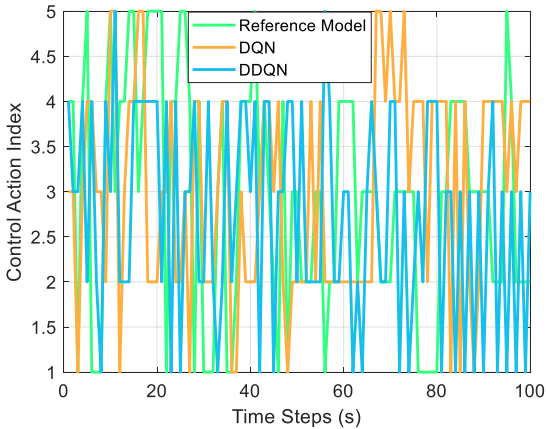

**FIGURE 8.** Control actions in one successful episode of three compared methods: Index=1, changing left lane; Index=2, idling speed; Index=3, changing right lane; Index=4, running faster; Index=5, running slower.

Furthermore, to defense the concrete control actions are different in these three methods. The curves of control action sequences of one successful episode (means the ego vehicle could drive from the starting to ending point) are given in Fig. 8. The actions in longitudinal and later directions of the ego vehicle are uninformed as five selections. They are changing left lane, changing right lane, idling speed, running faster, running slower. The differences between these trajectories indicate the proposed decision-making policy is different from the two benchmark methods (in the same successful episode). Overall, according to all the display results in this subsection, the optimality of the DDQN-enabled decision-making strategy is illuminated.

### B. COMPARISION BETWEEN DQN AND DDQN

Since DQN and DDQN are two permanent DRL algorithms, this experiment aims to appraise the learning and training procedure of these two approaches. As the target of the neural network is acquiring the mutable Q table. The normalized mean discrepancy of the Q table in the training process of these two methods is displayed in Fig. 9. The downtrend graphs indicate that both ego vehicles become more familiar with the driving environment by interacting with it. Furthermore, it can be discerned that the DDQN could learn more knowledge about the traffic situations with the same episodes, and thus results in the faster learning course. Hence, the ego vehicle could manipulate more efficiently and safely by the guidance of the DDQN algorithm.

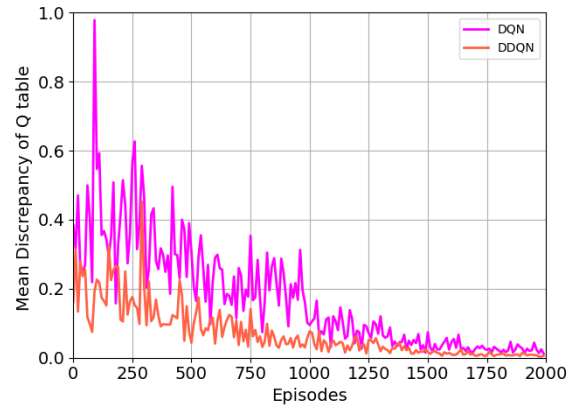

**FIGURE 9.** Mean discrepancy of Q table in the training process of two DRL approaches.

To exhibit the usage of the dueling network in future decisions in this autonomous driving problem, Fig. 10 discusses the track of the cumulative rewards. The uptrend variation implies that the control action choices are capable of improving future rewards. As the DDQN is larger than DQN, it signifies the related agent could achieve better control performance. This is also attributed to the advantage network in DDQN, which enables the ego vehicle to quantify the potential worth of current control action. To assess the above-mentioned decision-making policies in a similar driving condition, the next subsection discusses the adaptability of these strategies.

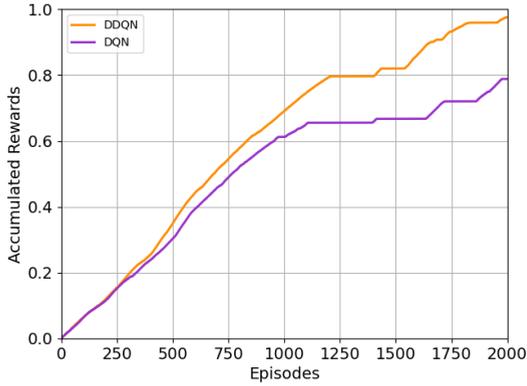

**FIGURE 10. Accumulated rewards in DQN and DDQN: the higher accumulated reward indicates better control action choices.**

### C. ADAPTABILITY ESTIMATION

After learning and training the automated vehicles in highway driving environments, a short episode is applied to test their adaptive capacity. The testing number of episodes is 10 in this work. The default settings and the number of lanes and surrounding vehicles are the same as the training process. The learned parameters of the neural networks are saved and can be utilized directly in the new conditions. The most concerning elements are the average reward and collision conditions of the testing operation.

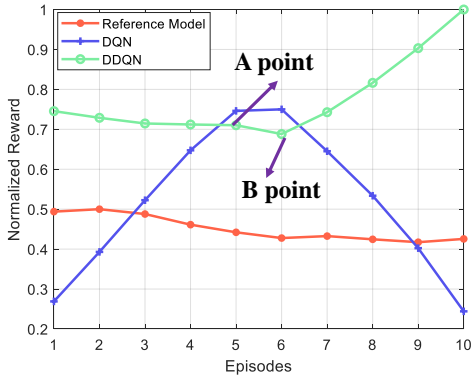

**FIGURE 11. Normalized reward in the testing experiment of three compared methods.**

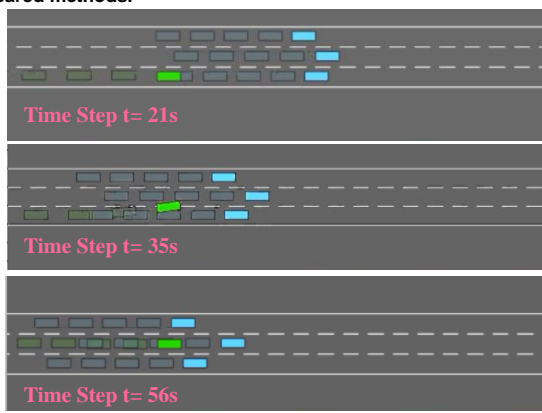

**FIGURE 12. One typical testing driving condition: the ego vehicle has to execute car-following behavior for a long time.**

Fig. 11 shows the normalized average reward of the reference model, DQN, and DDQN methods in the testing experiment. From (27), the reward is mainly influenced by the collision conditions and vehicle speed. The average reward may not achieve the highest score (100 in this work) because the ego vehicle has to slow down sometimes to avoid a collision. The ego vehicle also needs to change to other lanes to realize the overtaking process. Without loss of generality, two typical situations (two episodes, A and B points in Fig. 10) are chosen to analyze the decision-making behaviors of the ego vehicle.

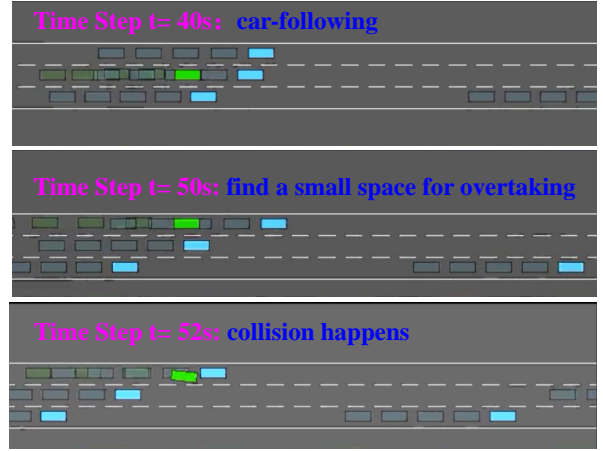

**FIGURE 13. Another representative testing driving condition: the ego vehicle make a dangerous lane changing and a collision happens.**

Fig. 12 depicts one driving situation that there are three surrounding vehicles in front of the ego vehicle (the episode represented by A point). The ego vehicle has to execute the car-following maneuver for a long time and wait for the opportunity the overtake them. As a consequence, the vehicle speed may not reach the maximum value, and the ego vehicle may not surpass all the surrounding vehicles before the destination. Furthermore, an infrequent driving condition is described in Fig. 13 (the episode represented by B point). The ego vehicle wants to achieve a risky lane-changing to obtain higher rewards. However, it cashed nearby vehicles because the operation space is not enough. This situation may not happen in the training process, and thus the ego vehicle could cause a collision.

TABLE I
TRAINING AND TESTING TIME OF DQN AND DDQN METHODS

| Techniques[#] | Training Time (h) | Testing Time (s)[*] |
|---|---|---|
| DQN | 8 | 27.89 |
| DDQN | 6 | 26.56 |

[#] A 2.30 GHz microprocessor with 31.8 GB RAM was used.
[*] The time that uses the trained parameters in a new driving situation.

Based on the detailed analysis in Fig 12 and 13, it hints us to spend more time to train the mutable decision-making strategy. These results also remind us that the relevant control policy has the potential to be applied in real-world environments. Table I provides the training and testing time of the DQN, and DDQN approaches. Although the training time can be only realized offline, the learned parameters and policies are able to be utilized online. This inspires us to implant our decision-making policy in the visualization simulation environments and to conduct the related loop experiments in the future.

## V. CONCLUSION

This paper discusses the highway decision-making problem using the DRL technique. By applying the DDQN algorithm in the designed driving environments, an efficient and safe control framework is constructed. Depending on a series of simulation experiments, the optimality, convergence rate, and adaptability are demonstrated. In addition, the testing results are analyzed, and the potentials of the presented method to be applied in real-world environments are proven. Future work includes the online applications of highway decision-making by executing hardware-in-loop (HIL) experiments. Moreover, the real-world collected highway database can be used to estimate the related overtaking strategy.